# De tauroctonías y estrellas: Mitra y la vida de una imagen

# On tauroctonies and stars: Mithras and the life of an image

Alejandro Gangui[1]

**Resumen**
Un héroe o un dios, con atuendo oriental y gorro frigio, apoya su cuerpo sobre el de un toro para someterlo. Con su rodilla flexionada presiona el lomo del animal y sujeta con la mano izquierda su cabeza a la vez que con su diestra clava un puñal en el cuello de la bestia. Refiriéndose a esta representación del sacrificio del toro, o tauroctonía, Fritz Saxl escribió una vez que no imaginaba otro caso en el que se pudiese observar con tal precisión el nacimiento de una imagen. Aunque la iconografía completa del culto a Mitra varía substancialmente de un santuario a otro, la tauroctonía, elemento siempre presente en todos los templos, se considera clave para su ideología. Aby Warburg no dejó de lado esta imagen, y la reprodujo en abundancia en el *Panel 8* sobre la Ascensión hacia el Sol de su *Bilderatlas Mnemosyne*. En este trabajo proponemos realizar un recorrido por la vida de esta imagen que, muy tempranamente, fue relacionada con las constelaciones, con los grandes círculos de la esfera celeste y con ciertos momentos singulares del ciclo estacional. Concluiremos con un análisis crítico de recientes especulaciones que ponen en juego aspectos sutiles del movimiento del cielo, impulsado por un dios poderoso que residía más allá de las estrellas, y los relacionan con la muerte simbólica del toro.

**Palabras clave**
Culto a Mitra. Iconología y Astronomía. La esfera celeste. Constelaciones de Orión y de Perseo. Aby Warburg.

**Abstract**
A hero or a god, wearing an oriental garb and a Phrygian cap, presses his body on that of a bull to subdue it. With his flexed knee he presses the back of the animal and holds its head with his left hand, while with his right hand he stabs a dagger into the neck of the beast. Referring to this representation of the sacrifice of the bull, or tauroctony, Fritz Saxl once wrote that he could not imagine another case in which the birth of an image could be observed with such precision. Although the complete iconography of the Mithras cult varies substantially from one sanctuary to another, the tauroctony, an element always present in all temples, is considered key to its ideology. This image was not neglected by Aby Warburg and was reproduced in abundance in *Panel 8* on the Ascension towards the Sun in his *Bilderatlas Mnemosyne*. In this work we propose to make a journey through the life of this image that, very early on, was related to the constellations, to the great circles of the celestial sphere and to certain singular moments of the seasonal cycle. We will conclude with a critical analysis of recent speculations that bring into play subtle aspects of the movement of the sky, driven by a powerful god who resided beyond the stars, and relate them to the symbolic death of the bull.

**Keywords**
Mithras cult. Iconology and Astronomy. The celestial sphere. Orion and Perseus constellations. Aby Warburg.

---

[1] Universidad de Buenos Aires, Facultad de Ciencias Exactas y Naturales, Argentina. CONICET – Universidad de Buenos Aires, Instituto de Astronomía y Física del Espacio (IAFE), Argentina.

De origen incierto, quizá procedente de Persia según los primeros investigadores del siglo XIX, el culto a Mitra, dios que somete al toro, floreció en el Imperio romano a partir del siglo I d.C. por el contacto de tropas y comerciantes con los territorios conquistados. A partir de entonces sus santuarios, en general subterráneos o ubicados en grutas naturales, se multiplicaron por toda Europa y el Mediterráneo, manteniéndose siempre como una religión mistérica cuyos ritos y doctrinas no dejaron registros escritos.

La representación aparecía indefectiblemente en el lugar más prominente de los templos, lo que ha llevado a los investigadores a tratar de descifrar su significado, original o heredado en el contexto de cada época, con la esperanza de comprender más sobre los aspectos internos de este culto mistérico que, como sabemos, llegó a Occidente y se constituyó en un importante rival del cristianismo en los primeros siglos después de Cristo.

Por otra parte, y tiempo antes de que se definiera la estructura presente del zodíaco en Babilonia, la observación atenta del cielo fue una ocupación que se concentraba esencialmente en ciertas estrellas brillantes, especialmente en aquellas ubicadas cerca del camino anual del Sol a lo largo de la eclíptica[2]. En aquella época, que para fijar ideas ubicaremos en los años de los agricultores del Creciente Fértil (levante Mediterráneo, Mesopotamia y Persia), varias de estas estrellas tenían apariciones llamativas en momentos previos al amanecer (ortos helíacos), lo que las convertía en marcadores privilegiados en determinadas fechas de relevancia calendárica.

Las estrellas de Tauro, por ejemplo, y en particular el grupo de las Pléyades, cuando coincidían en el cielo con la posición del Sol, virtualmente desaparecían del alcance de los observadores por lo menos por unos 40 días. Pero el desplazamiento del Sol sobre el fondo de las estrellas hacía que, poco más de un mes más tarde, estos notables grupos de estrellas reaparecieran al alba por el oriente antes que el Sol, y ese orto helíaco de las estrellas, allá por el año 4000 a.C., coincidía con el equinoccio vernal (nuestro actual equinoccio de marzo). Algo similar sucedía con la lúcida de Leo, la estrella Regulus, pero en una fecha cercana al solsticio de verano boreal. Y en lo que luego se conoció como Escorpio, la estrella Antares, ubicada en el corazón del escorpión, tenía su orto helíaco acompañada del equinoccio de otoño boreal.

---

[2] La eclíptica es la trayectoria curva "aparente" que recorre el Sol en el cielo durante todo un año vista por un observador terrestre. Sobre esa línea se ubican las constelaciones zodiacales.

Así, unos 6000 años atrás, estas estrellas marcaban momentos relevantes en el año, y seguramente fueron tomadas en cuenta por observadores de latitudes moderadas del hemisferio norte, de alrededor de los 30º de latitud, como en la ciudad sumeria de Ur, fundada alrededor del 3000 a.C. sobre el río Éufrates. Hay registros de asentamientos agrícolas con sistemas de irrigación en cercanías de Ur que datan de más de 7000 años. Las condiciones necesarias para la práctica de la astronomía estacional, entonces, estaban dadas con seguridad para el año 4000 a.C. cuando, como mencionamos más arriba, las estrellas prominentes de tres constelaciones sumerias (Escorpio, Leo y Tauro) dividían el calendario agrícola y ritual.

**El motivo del combate entre el León y el Toro**

Algunas de estas figuras del cielo, sin embargo, no permanecían en paz, o al menos no lo hacían así en la mente de los antiguos observadores. Hartner (1965) señala que las representaciones del combate entre el León y el Toro se han venido repitiendo por varios miles de años. La lucha de la pareja, que hoy podemos encontrar en las esculturas monumentales que decoran los palacios aqueménidos de Persépolis, es uno de los motivos más antiguos y, sin duda, el rastro más tenaz de la historia del arte del Cercano Oriente. Para demostrarlo, Hartner compara esta escena en varios soportes: en un sello Elamita del cuarto milenio a.C., en la decoración de una copa Sumeria del tercer milenio a.C., en una gran pared esculpida de la ya mencionada capital del imperio persa del siglo VI a.C., y finalmente, en una miniatura persa (manuscrito iluminado) del período mogol de aproximadamente el siglo XIII de nuestra era [fig. 1].

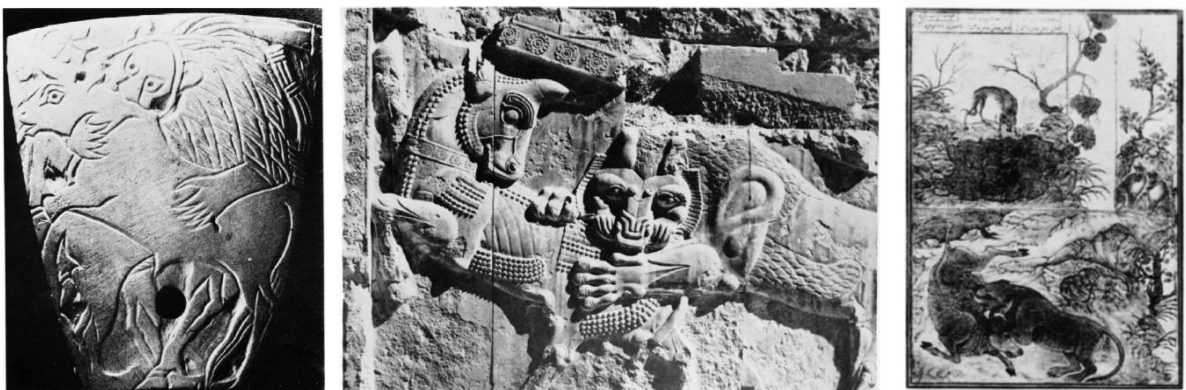

Fig. 1. Representaciones del combate entre el León y el Toro en una copa Sumeria del tercer milenio a.C. (imagen izquierda), en una pared esculpida de Persépolis del siglo VI a.C. (centro) y en un manuscrito iluminado persa del período mogol (derecha) (Hartner, 1965: Plates II, I, III, respectivamente).

Evidentemente, dice Hartner,

> a lo largo de cinco mil años, a pesar de innumerables guerras, de la caída de docenas de viejos estados o reinos y del ascenso de otros tantos, asistidos por cambios de las razas e idiomas dominantes, no sólo la combinación simbólica de los dos animales combatientes, sino incluso algunos detalles muy característicos en el modo de representación fueron cuidadosamente conservados. Este solo hecho merece nuestra atención y justifica el interés que los arqueólogos y los historiadores del arte modernos le han dedicado a este motivo (Hartner, 1965: 1).

Y teniendo en cuenta la permanencia que la imagen de la muerte del toro antiguo ha tenido en el arte posterior, podríamos preguntarnos: ¿es este un caso representativo de "la vuelta a la vida de lo antiguo" (*das Nachleben der Antike*, en las palabras de Warburg)? Claro que aquí no nos detendremos en el descubrimiento, por ejemplo, de la ninfa como Leitmotiv perenne de la evocación vivaz del paganismo, con sus gráciles movimientos de miembros, cabellos y vestimentas (de hecho, estos no aparecen aquí), ni llegaremos hasta el *Quattrocento* florentino, como le interesaba a Warburg (Burucúa, 2003: 15). Nuestro trabajo, más modesto, se limita a no más allá de los primeros siglos de nuestra era.

Ahora bien, proyectando la escena del combate en el cielo de hace 6000 años, la imagen cobra sentido [fig. 2]. El león mata al toro unos 40 días antes del equinoccio de primavera boreal de aquella época, en los días en que se inicia el arado y la siembra (comenzando el ciclo agrícola). En esos días, la constelación del toro (o al menos Aldebarán y otras de sus estrellas brillantes, e incluso las Pléyades) se oculta en el occidente poco después de la puesta del Sol. El toro sigue al Sol en su viaje hacia la tierra de los muertos. En ese mismo instante, al comienzo de la noche, Leo se halla alto en el cielo hacia el sur y reina sobre las estrellas. Si bien esta escena parece repetirse con el correr de los días, la "muerte" del toro se hace evidente porque el Sol se desplaza en el cielo hacia su conjunción con Tauro. Extraviado en la intensa luz solar, el toro desaparece del cielo del crepúsculo vespertino, como una víctima del león quien, victorioso, aún reina en lo alto del cielo del inicio de la noche.

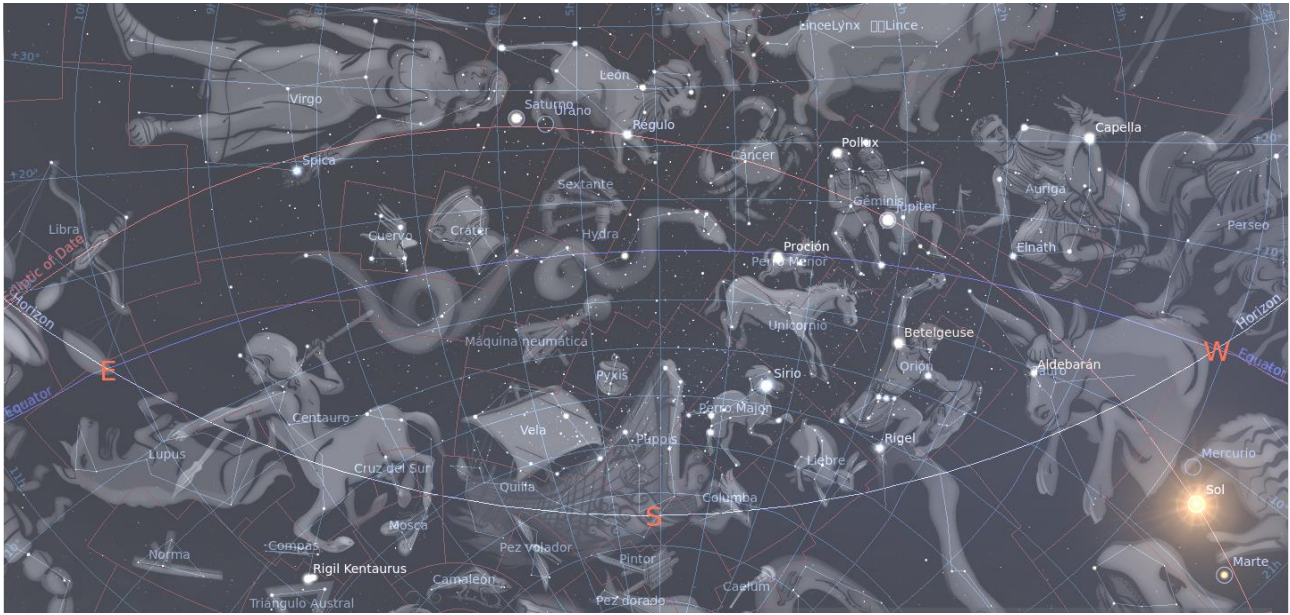

Fig. 2. Aspecto del cielo desde una latitud aproximada de 35º N, en el año 4000 a.C. El punto equinoccial se halla sobre los cuernos del toro y al Sol, ya oculto en el horizonte occidental, le llevará unos 40 días más alcanzarlo, iniciando de esa manera la primavera boreal. La imagen es sólo indicativa; muchas de las constelaciones mostradas son sólo una guía para el lector, y no corresponden a las conocidas en esa época. Imagen del autor a partir de un mapa cortesía de *Stellarium*.

Pero unos 40 días más tarde, el toro reaparece en el cielo, esta vez en el crepúsculo del alba, en lo que ya llamamos el orto helíaco de Tauro. La fecha coincide con el equinoccio vernal (primavera boreal) que, tradicionalmente, era el momento del renacimiento de la vida en el Cercano Oriente antiguo [fig. 3].

Sin embargo, los siglos pasaron y la precesión de los equinoccios[3] fue alterando el cielo, o al menos la ubicación de los puntos equinocciales en el cielo de las constelaciones. De hecho, fue gracias a Ptolomeo, quien lo consignó en su *Almagesto*, VII: 2, que las observaciones de Hiparco sobre la precesión pudieron llegar hasta nosotros. En el tratado *Sobre la duración del año*, hoy perdido, Hiparco había escrito que solsticios y equinoccios se corrían hacia el oeste con respecto a las estrellas "no menos de 1/100º" en un año. Hoy sabemos que este valor es algo bajo (el valor más preciso es 1º en 72 años), pero lo importante es que este sutil movimiento del cielo (en aquel momento de origen incierto) ya había sido develado en el siglo II a.C., y es virtualmente la única parte de la maquinaria griega del cielo que aún hoy se considera "correcta" (Evans, 1998: 262).

---

[3] Responsable de este cambio es el "bamboleo" del eje de rotación de la Tierra, el cual hace que todas las estrellas del cielo nos parezcan irse desplazando año tras año de sus posiciones. Entre las estrellas se cuentan, por supuesto, las que indican los puntos equinocciales, es decir los lugares del cielo que, vistos desde la Tierra, coinciden con la posición del Sol en el comienzo de la primavera y del otoño, o, como es más frecuente decirlo, los lugares por los que 'pasa' el Sol en esos momentos. Este es el origen de la expresión *precesión de los equinoccios*.

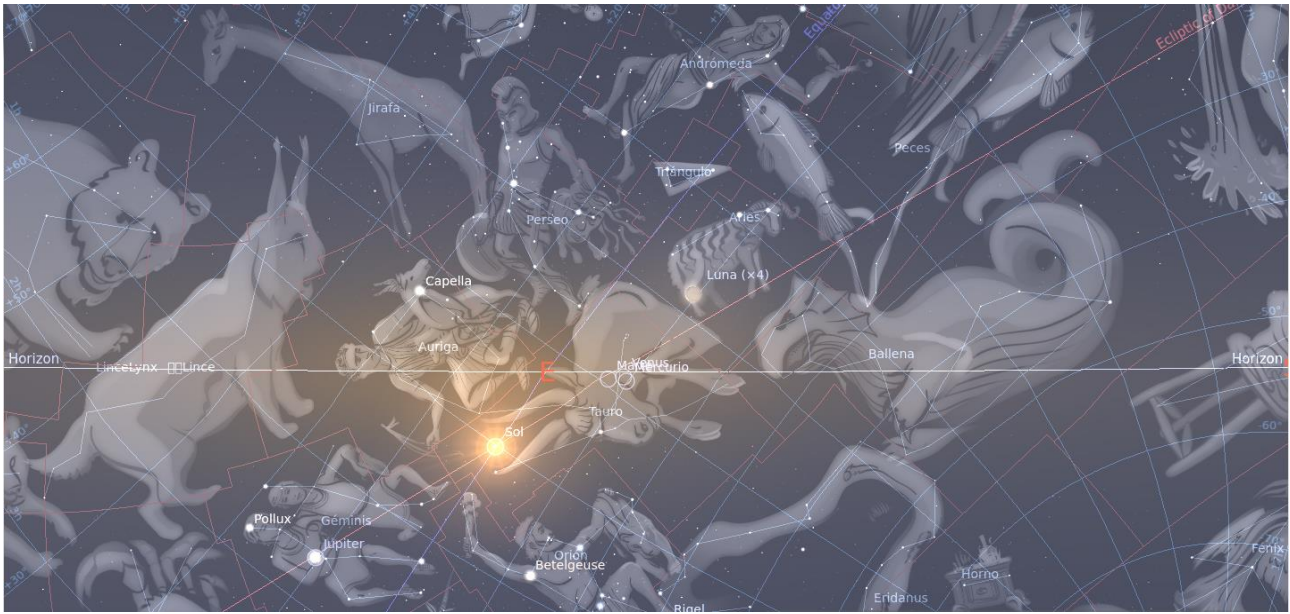

Fig. 3. Aspecto del cielo del oriente antes del amanecer con el Sol ubicado en el equinoccio vernal. Corresponde a la primera aparición de algunas estrellas brillantes de Tauro (su orto helíaco) antes de que el Sol naciente las oculte con su brillo. Las demás características son idénticas a las de la figura anterior. Imagen del autor a partir de un mapa cortesía de *Stellarium*.

Volviendo a nuestro tema, y a pesar de la alteración del cielo debida a la precesión, para el siglo VI a.C. la dinastía aqueménida de los reyes persas todavía daba un lugar de relevancia al motivo del combate entre el León y el Toro (como vimos en la imagen de la gran pared esculpida de Persépolis). Pero si nos ubicamos unos 1000 años antes, hacia el año 1500 a.C., vemos que el punto equinoccial ya se ha corrido unos 35º desde los cuernos hacia las patas del toro, y por ello Tauro desaparece del cielo ("quemado" por el Sol) en una fecha más cercana al equinoccio vernal. Así, en esa época la muerte crepuscular del toro coincidía con el festival equinoccial del año nuevo del imperio persa, y un mito clave del culto de Zoroastro en esas tierras involucraba justamente la muerte del Toro de la Creación como presa de Ahrimán (llamado Angra Mainyu, en el *Avesta*), el Dios del Mal. De acuerdo con Edwin C. Krupp (1992), la versión persa del toro cósmico y la importancia del dios Mitra, que en el zoroastrismo tenía el rol de mediador, pero en Persia era un dios del cielo y la luz heredado de la tradición védica, llevó a Franz Cumont (1896-1899), uno de los iniciadores del Mitraísmo, a asociar los conocidos sacrificios del toro que abundaban en los santuarios del Imperio romano con la antigua doctrina iraní. Vale la pena aclarar, como lo ha enfatizado Roger Beck (2015), que fue el dios (Mitra) y no su culto, lo que fue importado en el Imperio romano desde Persia (el origen del Mitraísmo es aún dudoso). Pero no nos adelantemos, y veamos cómo fueron los orígenes de la doma del toro -no ya su sacrificio por parte de un león-, la victoria del humano sobre la poderosa bestia que se convirtió en la imagen del poder heroico de ahí en más.

**La victoria del humano sobre la bestia: la doma del toro**

Fritz Saxl (1989) se encargó de hacer la conexión desde las primeras imágenes del "combate" entre el toro y los héroes de turno, hasta su sacrificio final característico de la iconografía del Mitraísmo romano. Señaló unos sellos acadios donde la estilización es poco usual, dado que el héroe muestra una altura similar a la de su presa [fig. 4]. Puede aparecer de pie o inclinado, o puede incluso sujetar al enorme toro por la cola, volviéndolo indefenso, pero siempre los combatientes tienen tamaño similar.

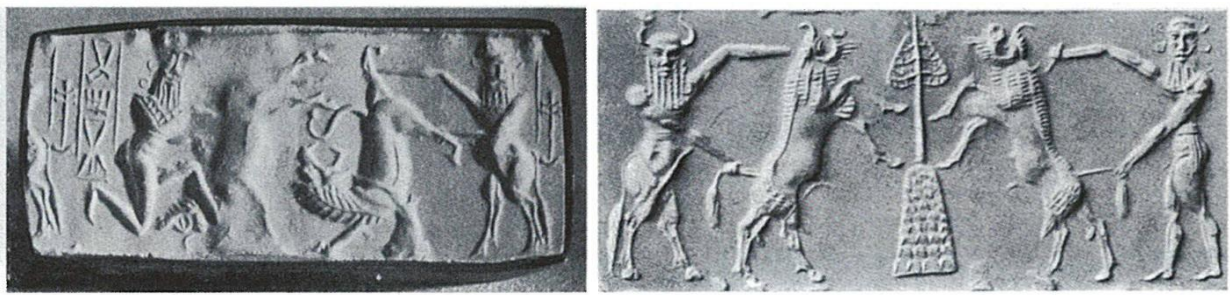

Fig. 4. Sellos acadios representando un hombre y un toro (París, Bibliothèque National; imagen izquierda) y un héroe y un hombre toro dominando a toros (Londres, British Museum; imagen derecha), ambos de aproximadamente el II milenio a.C. (Saxl, 1989: 367).

Otras imágenes muestran a un hombre dominando al toro en Micenas, "rica en oro", el reino del héroe homérico Agamenón de la Guerra de Troya en la Grecia arcaica, pero en esta representación el luchador está arrodillado y mantiene la cabeza de la bestia hacia abajo. Esto le recuerda a Saxl la imagen de Hércules derrotando al centauro Neso, ya en tiempos clásicos. Nuevamente aquí, el héroe aparece de rodillas junto al monstruo derrotado [fig. 5].

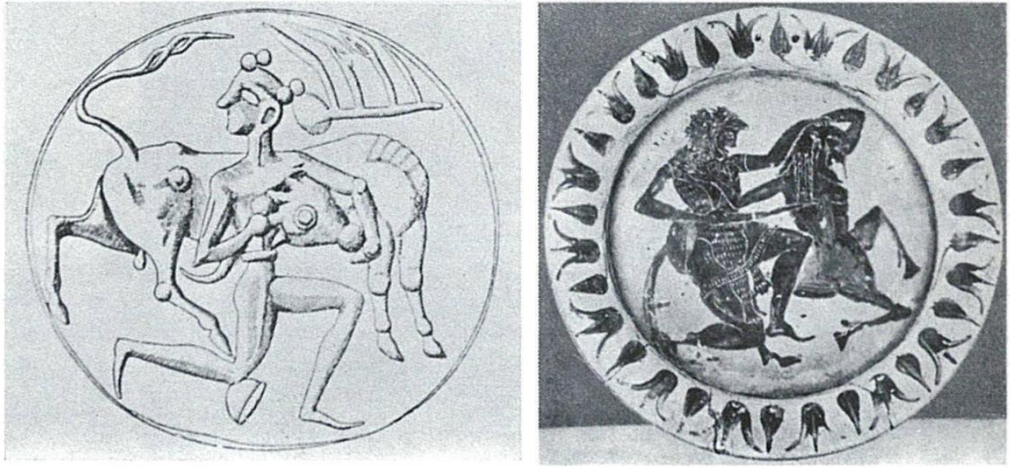

Fig. 5. Gema grabada de Micenas que representa a un hombre que subyuga a un toro (Atenas, Museo Nacional; imagen izquierda) y disco de figuras negras que muestra a Hércules derrotando a Neso (Munich, Museum Antiker Kleinkunst; imagen derecha) (Saxl, 1989: 367).

Es entonces que, señala Saxl (1989: 14), "estamos ahora cerca del clímax". Entre los siglos VI y V a.C. en las importantes escuelas de escultura griega arcaica, surgió una nueva imagen que se hizo decisiva. Esta muestra al hombre con una de sus rodillas sobre el animal y con el pie de la otra pierna afianzado sobre el suelo, a la vez que aferra con una de sus manos la cabeza de la bestia. "Esta imagen se hizo clásica en el momento en que se inventó", dice el investigador, y cita varios ejemplos donde la representación toma vida: alguna hazaña del ya mencionado Hércules, por supuesto, pero también, y más asombrosamente, la escena de Belerofonte domando a Pegaso, el caballo volador [fig. 6], o la lucha contra las amazonas, donde "el conquistador pone la rodilla sobre la amazona exactamente de la misma manera que Hércules la pone sobre el animal" (Saxl, 1989: 15).

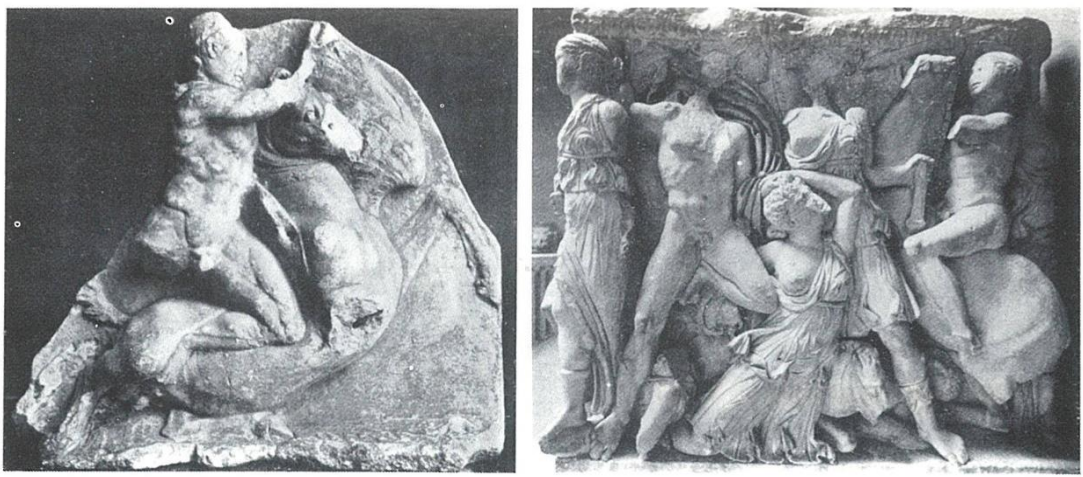

Fig. 6. Belerofonte domando a Pegaso (Budapest, Museum; imagen de la izquierda) y relieve de un sarcófago con Amazonas (Londres, British Museum; a la derecha) (Saxl, 1989: 369).

Se podría decir que es ahora una imagen que ha alcanzado la perfección: todos los intentos anteriores de representación de la lucha entre el hombre y la bestia caen en el olvido, y sólo queda este. Tal fue su ubicuidad con fines decorativos que perdió su carácter impresionante. Sin embargo, afirma Saxl, llegó un momento en el que, tras siglos de existencia, la imagen asumió una importancia capital: "Fue el momento en que las divinidades griegas y romanas perdieron su influencia, y se crearon nuevos cultos basados en el impacto de las religiones orientales" (Saxl, 1989: 15). Es aquí que nuestra historia nos lleva a Mitra.

**El culto oriental a Mitra**

Como ya señalamos, Mitra era un dios del cielo y de la luz, aunque también del comercio, heredado por los primeros pueblos persas de la tradición védica. Su nombre aparece mencionado en los libros del *Avesta*, la colección de textos sagrados zoroástricos de la antigua Persia, y en los primitivos textos indios. Seguiremos aquí de cerca el trabajo de Romero Mayorga (2016), quien ha hecho un detallado rastreo bibliográfico para su reciente tesis sobre la Iconografía mitraica en Hispania.

En sus orígenes Mitra era una divinidad indoeuropea que fue considerada el dios de los acuerdos, pues su nombre en el dialecto del *Avesta* revela el concepto ético del buen comportamiento social, y significa tratado o contrato. Las creencias religiosas de las tribus de lengua iraní instaladas en Turquestán occidental entre los milenos II-I a.C. estaban íntimamente vinculadas con aquellas practicadas por los indoarios, especialmente a partir del año 1700 a.C. Mitra está presente en ambos credos como una divinidad solar. Se piensa que en la vertiente védica, Mitra se escindió en otras dos divinidades: Ariamán y Varuna.

La raíz "Mitr-" aparece por primera vez en una tablilla de Bogazköy (la antigua Hattusa, capital del Imperio hitita), junto a Varuna, como garante de un tratado firmado por los reyes Suppiluliuma de Hattusa y Mitavaza de Mitani en el año 1380 a.C. Por otra parte, en el *Rig veda*, el más antiguo de los textos religiosos de la literatura india, compuesto oralmente en sánscrito a mediados del II milenio a.C., Mitra es el dios encargado de mantener el orden cósmico y velar por la buena conducta religiosa y moral.

Volviendo a la antigua Persia, Mitra es una divinidad que a través del fuego ejerce la ley, recibe sacrificios sangrientos y su gran fiesta se celebra en el equinoccio de otoño, cuando se lo identifica con el Sol que todo lo ve. Aquí es donde hallamos la evidencia más antigua que liga a Mitra con el Sol, durante los siglos VI-V a.C.: en un himno del *Avesta* se lo exalta como el dios que tiene diez mil espías ("Es vigilante y nunca duerme", dice Saxl (1989: 26), "tiene mil oídos para oir y diez mil ojos para ver y proteger a sus fieles"). Es fuerte, omnisciente y posee un carro tirado por cuatro blancos corceles y una de sus ruedas es de oro, una clara alusión a la cuadriga solar (Saxl, 1989: 25). En el dualismo zoroástrico, Mitra es, junto a Ahura Mazda, la luz en combate permanente con la oscuridad y quien ahuyenta a los malos espíritus, encarnados en Ahrimán. Es quizás en este momento cuando recibe su epíteto más habitual: *invictus* (Romero Mayorga, 2016: 120).

Ahora bien, desde el período más arcaico, a la ya mencionada cualidad moral del carácter de Mitra (como dios de la confianza y de los acuerdos) se corresponde también un aspecto natural. Como dios del cielo, Mitra es asimismo dador de fertilidad y de riqueza (Saxl, 1989: 21). Mitra-Varuna se llaman los grandes toros, y en el *Avesta*, Mitra es la gran divinidad protectora del ganado. Sin embargo, en la religión mística de las legiones romanas que invadieron Europa en el siglo I a.C., Mitra se convirtió en un héroe que mata a un toro. Pues esta es la imagen prominente que se destaca en los mitreos y que es la principal fuente de información para los estudios iconográficos, en un área donde no han quedado textos originales sobre este culto en el que los fieles habían jurado mantener el silencio. Y Saxl se pregunta: "¿Cómo se convirtió Mitra, el poderoso toro, el que protege los rebaños, en una divinidad destructora de toros?". Si bien en el campo de estudios dedicado a los misterios de Mitra existe una línea de investigación que busca establecer la continuidad del culto indoiranio con el romano, esta continuidad resulta muy difícil de localizar.

**El Mitraísmo en el Imperio romano**

La nueva imagen del dios oriental, por supuesto, recibió un significado completamente distinto en su renacimiento en occidente. El Mitra romano ya no protege a los rebaños. En cambio, se alía con sus antepasados – que Saxl nos mostró en los sellos mesopotámicos – y mata al toro en la caverna del mundo para producir fertilidad.

Se piensa que el nuevo culto mistérico surgió en el siglo I a.C. en Cilicia, en la costa meridional de Anatolia (hoy sudeste de Turquía) en Asia Menor, y la fuente escrita más antigua que lo refiere es la obra de Plutarco. En su colección biográfica *Vidas paralelas*, el historiador griego cuenta la introducción del culto en el Imperio romano a través de los piratas cilicios capturados por Pompeyo (106 - 48 a.C.), quien en el año 67 a.C. fue el encargado de limpiar el Mediterráneo de su acecho, restaurando la comunicación marítima entre Hispania, África e Italia. Cuenta Plutarco que los piratas se habían atrevido a saquear los templos y que también hacían "sacrificios traídos de fuera, como los de Olimpia, y celebraban ciertos misterios indivulgables, de los cuales todavía se conservan hoy en el de Mitra, enseñado primero por aquellos" (Campos Méndez, 2010).

A pesar del texto de Plutarco y de considerar que aquellos piratas fueran efectivamente iniciados en los misterios de Mitra, es muy poco probable que el culto tuviera la gran difusión que conocemos a partir de un pequeño grupo de nómades marítimos fuera de la ley romana. Los restos arqueológicos más tempranos hallados en esta zona son de mediados del siglo II d.C., por lo que se considera que el Mitraísmo fue una religión activa en el Imperio romano desde los inicios del siglo II hasta finales del siglo IV (Beck, 2015).

El culto a Mitra fue muy popular entre los soldados y funcionarios menores, militares o civiles, como se refleja en los múltiples santuarios y objetos de culto que las excavaciones han ido descubriendo a lo largo y ancho del Imperio, desde el norte de Inglaterra hasta las costas del Mar Negro. Pero no llegó jamás a ser una religión de masas, ya sea en las grandes ciudades o en zonas rurales. Su atractivo atrajo a soldados, marinos, mercaderes y esclavos, quienes estaban preocupados por el destino de sus almas, ya que el propósito del culto era la salvación personal. Se piensa que el culto no involucraba a las mujeres, aunque estas ideas fueron más recientemente puestas en duda. Varios estudiosos han afirmado que la exclusión de la mujer del ámbito mitraico es una construcción moderna, ajena a los hallazgos arqueológicos más recientes (David, 2000; Griffith, 2006).

En Roma, los rituales y doctrinas secretos del Mitraísmo incorporaron metáforas celestiales para transmitir más eficazmente el mensaje de salvación personal. Es aquí que la imagen cobra vida renovada bajo la égida de la bóveda de las estrellas.

**La tauroctonía como un mapa estelar**

La tauroctonía es la iconografía principal del culto, análogo en su importancia mítica y ubicación dentro del mitreo a la representación de la crucifixión en las iglesias cristianas (Beck, 2015). La imagen sigue una composición estándar que muestra a Mitra sacrificando al toro en la entrada de una caverna. Pero este no es un toro cualquiera; es el toro de la Creación, y todas las plantas y animales del mundo surgen de su cuerpo una vez que el héroe lo apuñala. Unas espigas de trigo florecen de su cola y su sangre se convierte en el vino sagrado que los iniciados bebían durante los servicios religiosos [fig. 7].

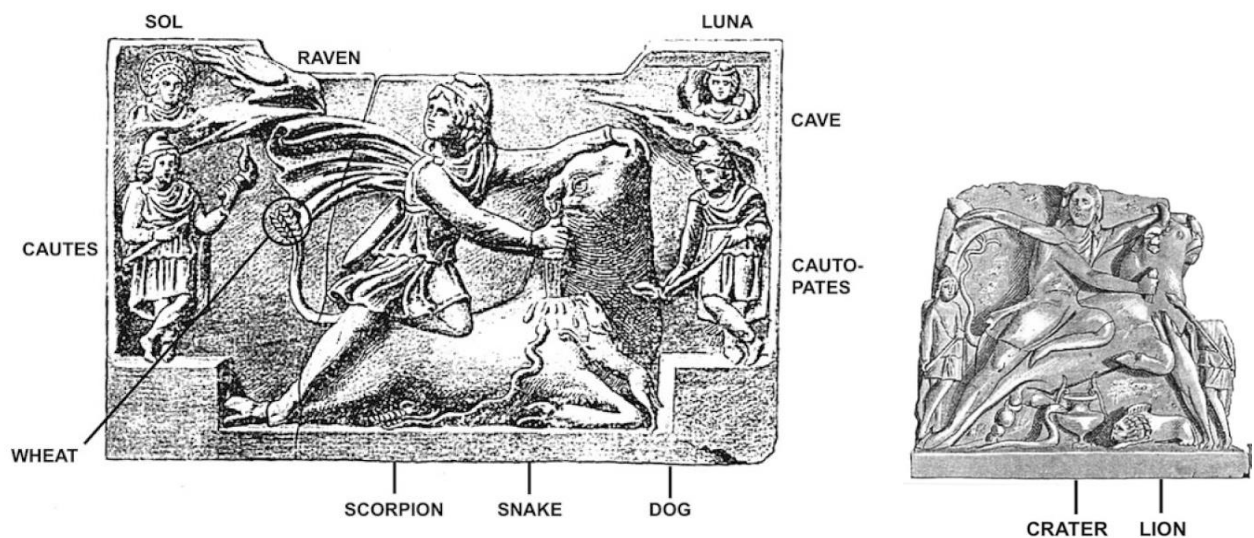

Fig. 7. Representación de una tauroctonía donde se señalan los detalles usuales de la composición iconográfica. Muchas veces también se incluyen una crátera y un león en la representación principal (Beck, 2015: 1674, 1675).

Las representaciones de Mitra en el acto de sacrificar al toro son notablemente consistentes en todos los santuarios de Europa. Al fondo de la "cueva", generalmente subterránea o ubicada en un nivel ligeramente inferior al del suelo, la imagen clave atrae la atención hacia el episodio principal del mito. La daga del héroe penetra en dirección al corazón de la bestia, al tiempo que varias espigas de trigo crecen de su rabo, mientras un perro y una serpiente aparentan organizar su propio ataque al toro debilitado, y un escorpión apunta sus pinzas hacia los genitales, claro símbolo de la fertilidad del toro. A veces algunas escenas muestran una crátera al centro, sobre el piso de la cueva (quizá donde se mezclaba el vino/sangre), y un cuervo y un león también asisten al sacrificio.

En numerosos altares en relieve están presentes los bustos de los dioses Sol y Luna en los ángulos superiores del campo compositivo, así como también dos personajes estantes, los dadóforos, muy similares a Mitra, que flanquean la escena principal mientras sostienen antorchas, identificados como Cautes y Cautopates. El primero sostiene una tea alzada y simboliza, entre varias posibles hipótesis, el Sol naciente. El joven Cautopates, en cambio, lleva una antorcha que apunta hacia abajo y simbolizaría el ocaso. A estos "gemelos" también se los llama Dioscuros, y se piensa que representan los opuestos luz/sombra, día/noche, amanecer/ocaso, primavera/otoño, etc., pero lo que hoy parece seguro es que el significado real de este par de personajes es tan incierto como el origen de sus nombres (ver, sin embargo, Beck, 2001: 70).

A fines de la década de 1860, K.B. Stark (1869) y otros (cf. Chapman-Rietschi, 1997) interpretaron estos emblemas de manera astronómica. Así, todos los personajes que aparecían en la tauroctonía tenían su imagen en el cielo, luego de que la constelación de Tauro, ya sumergida en el brillo del Sol, se ocultase por última vez (luego de su ocaso helíaco) en el horizonte occidental [fig. 8]. La oscuridad debajo de ese horizonte donde permanecía oculto el toro era la cueva donde se lo sacrificaba. Y mientras Tauro moría Leo dominaba el cielo y la constelación de Hydra, la serpiente del agua, se deslizaba por debajo del león. También se hacían presentes las constelaciones de Corvus, el cuervo, y Crater, la copa. Por su parte, el perro también estaba en el cielo, representado quizá por el Can Mayor o el Can Menor, o por la lúcida del primero, Sirio, mientras que Libra (la balanza), tradicionalmente relacionada con las pinzas del escorpión, en esos instantes subía por el horizonte oriental y acechaba al debilitado toro.

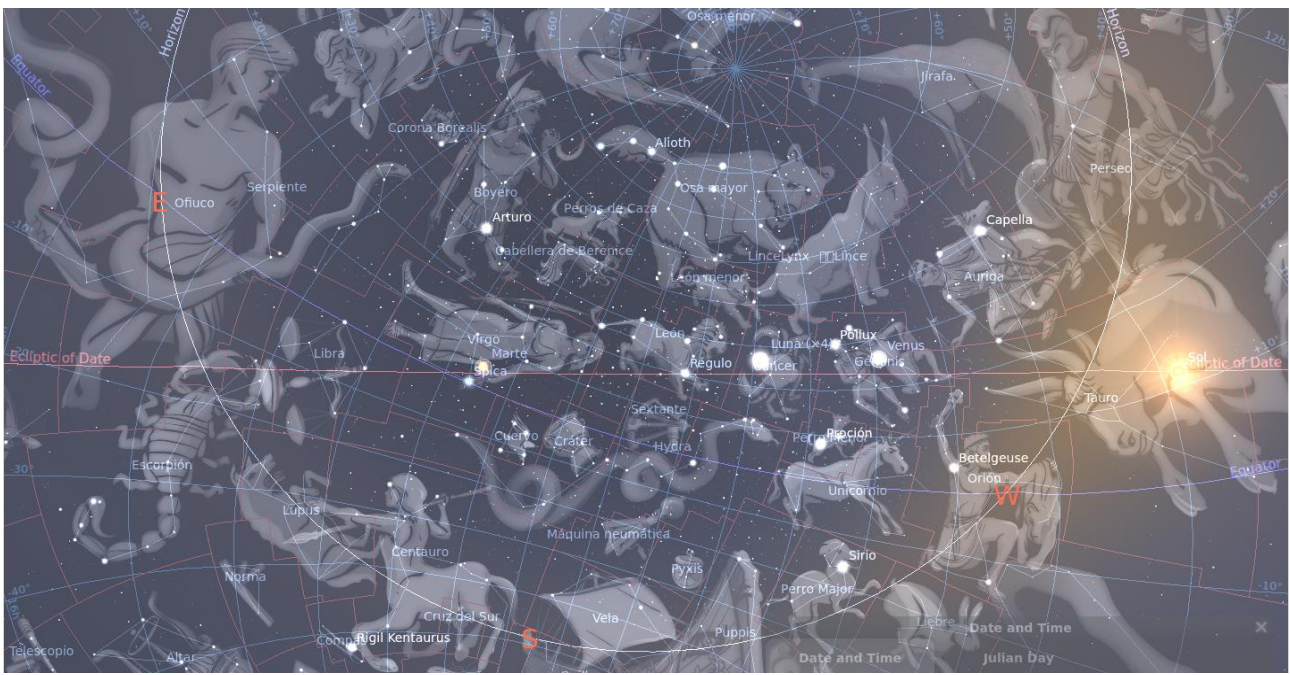

Fig. 8. Aspecto del cielo de la primavera (posterior al cruce del Sol por el equinoccio vernal, hacia la derecha y fuera de la imagen), visible desde la latitud de Roma en el siglo I a.C. El plano de la eclíptica, ubicación de los signos del zodiaco, se muestra horizontal por el medio de la imagen. En momentos del ocaso helíaco de Tauro, cuando el toro se oculta por el occidente, se llegan a ver todas las constelaciones que tienen sus representaciones en la tauroctonía. Imagen a partir de un mapa cortesía de *Stellarium*.[4]

Desde muy temprano en su estudio, entonces, se notó que los diferentes elementos en la composición correspondían simbólicamente a un grupo de constelaciones ubicado en el zodíaco y por debajo de él, desde Tauro en el occidente, hasta Escorpio en el otro extremo del cielo. La posibilidad de una mera coincidencia entre la iconografía y el cielo era muy difícil de admitir. Sin embargo, como enfatiza Beck (2015: 1675), hay una clara polisemia en la composición. Por ejemplo, los dadóforos gemelos no sólo representarían a los miembros de la constelación y signo de Géminis, sino también a todos los cuerpos celestes que surgen en el oriente y se ocultan en el occidente.

---

[4] Al igual que en las figuras anteriores, esta imagen del cielo es sólo indicativa; muchas de las constelaciones mostradas aquí son sólo una guía para el lector pues no corresponden a las conocidas en la época.

Entre otros misterios que quedan, no conocemos el momento preciso del año en el que Mitra sacrifica al toro y tampoco sabemos cuándo este acto mítico era recreado en los santuarios ocultos de los mitreos con el fin de bautizar a los iniciados con la sangre del animal sagrado. Sin embargo, la imagen de este sacrificio es enteramente consistente con el aspecto del cielo (en territorios romanos del siglo I a.C.), en una noche específica, poco después de la puesta del Sol, ya ocultándose el toro en el occidente durante la primavera, es decir, en los días posteriores al pasaje del Sol por el punto (equinoccial) vernal, cruce de los grandes círculos de la esfera de las estrellas: la eclíptica y el ecuador celeste.

**Tauroctonía y precesión de los equinoccios**

A pesar de este marco coherente – y quizás asombroso – en la interpretación astral de la imagen de la tauroctonía, no todos estaban convencidos. En particular, el autor de la primera gran obra dedicada a los estudios de Mitra, el ya mencionado Franz Cumont, aseguraba que el culto mistérico romano tenía su origen en el dualismo persa, y que el sacrificio del toro a manos del héroe formaba parte de la constante lucha entre el Bien y el Mal. Cumont se basaba en parte en la connotación negativa que animales como la serpiente, el escorpión y el perro tenían en el *Avesta*. Sin embargo, sus argumentos no eran del todo satisfactorios, pues asignaba la matanza del toro a Mitra y no a Ahrimán, que como vimos representaba el verdadero dios del Mal.

Pero hubo algo que quizá fue más grave: Cumont consideraba que los personajes de la tauroctonía no revestían relevancia astral, pues esas constelaciones no jugaban ningún rol en el mito zoroástrico. La estatura intelectual de este pionero congeló el estudio en el supuesto origen persa de Mitra hasta que, en la década de 1970, varios trabajos sugirieron que el simbolismo mitraico no debía su origen al antiguo Irán, sino que probablemente sería un culto desarrollado en el Mediterráneo. Más aún, se recuperó la idea de que el Mitraísmo se habría originado en Cilicia, precisamente como Plutarco lo había contado.

Ahora sí, como culto netamente occidental (mediterráneo), la representación de la tauroctonía como un mapa estelar debía completarse con el elemento quizá más importante: la identificación de la constelación que representaba al propio dios Mitra. No había inconvenientes sobre los paralelos entre el toro, el escorpión, la serpiente o el perro y sus constelaciones asociadas, pero con el héroe de la imagen aún quedaban dudas. Fue entonces que varios investigadores resucitaron viejas ideas astronómicas en la iconografía mitraica y, unos años más tarde, Michael Speidel sugirió que Mitra venía representado en la figura de Orión, el poderoso cazador que tradicionalmente combate contra el toro. Speidel (1980) enfatizaba la relevancia de las constelaciones ubicadas sobre el ecuador celeste y, en siglo I a.C., Orión era una de ellas (y lo es aún hoy, pero no lo era en épocas muy anteriores, debido a la precesión de los equinoccios). Esta identificación, por supuesto, era bien transparente en las estrellas, pero lamentablemente no arrojaba luz sobre la teología mitraica.

Un intento posterior de identificar a Mitra en el cielo se lo debemos a David Ulansey (1989: 26), quien sugirió que sería Perseo – el héroe mitológico célebre por haber seccionado la cabeza de la terrible Medusa – el encargado de sacrificar al toro. Su argumento une varios elementos que fueron ganando aceptación entre los estudiosos. De acuerdo con su teoría (y en acuerdo con los ya mencionados escritos de Plutarco), habría sido la ciudad de Tarso, capital de la provincia de Cilicia, la cuna en donde se originó la doctrina mitraica de la trascendencia cósmica, pero – y este es el elemento clave de la propuesta de Ulansey – como respuesta al descubrimiento de Hiparco del movimiento de precesión de los equinoccios en el año 128 a.C.

Vayamos por partes. *Primo*, sabemos que Tarso era un importante centro intelectual en filosofía estoica del siglo I a.C. Para el estoicismo, de clara orientación astral, el Cosmos era un ser vivo y complejo, y estaba gobernado por una divinidad invisible y poderosa que residía más allá de las estrellas y que movía el universo de acuerdo con su voluntad. Ulansey (1989: 43) sugiere que, en Tarso, esta divinidad había sido identificada con Perseo, el fundador legendario y héroe tutelar divino de la ciudad.

*Secundo*, recordemos que el movimiento de precesión, aunque extremadamente sutil, es detectable como un desplazamiento de los puntos equinocciales, es decir, de los lugares en la bóveda estelar en donde se cruzan la eclíptica y el ecuador celeste. El descubrimiento de Hiparco dejaba en claro que, antes del período greco-romano en el cual el punto vernal (o equinoccio de primavera boreal) se hallaba en Aries (el Carnero), la última constelación en la cual este equinoccio había ocurrido era en Tauro, que es su vecina.

Ese conocimiento de que el equinoccio vernal (uno de los momentos clave en el recorrido del Sol sobre la eclíptica) se había desplazado fuera de Tauro habría sugerido a los intelectuales de Tarso que la "Era del Toro" había llegado a su fin. Y materializaron este fenómeno astronómico con la muerte del toro de la caverna a manos de su héroe, Perseo. En este contexto, la ubicación casual de la constelación de Perseo justo por encima de la de Tauro, no extrañamente habría sugerido la imagen principal de la tauroctonía [fig. 9].

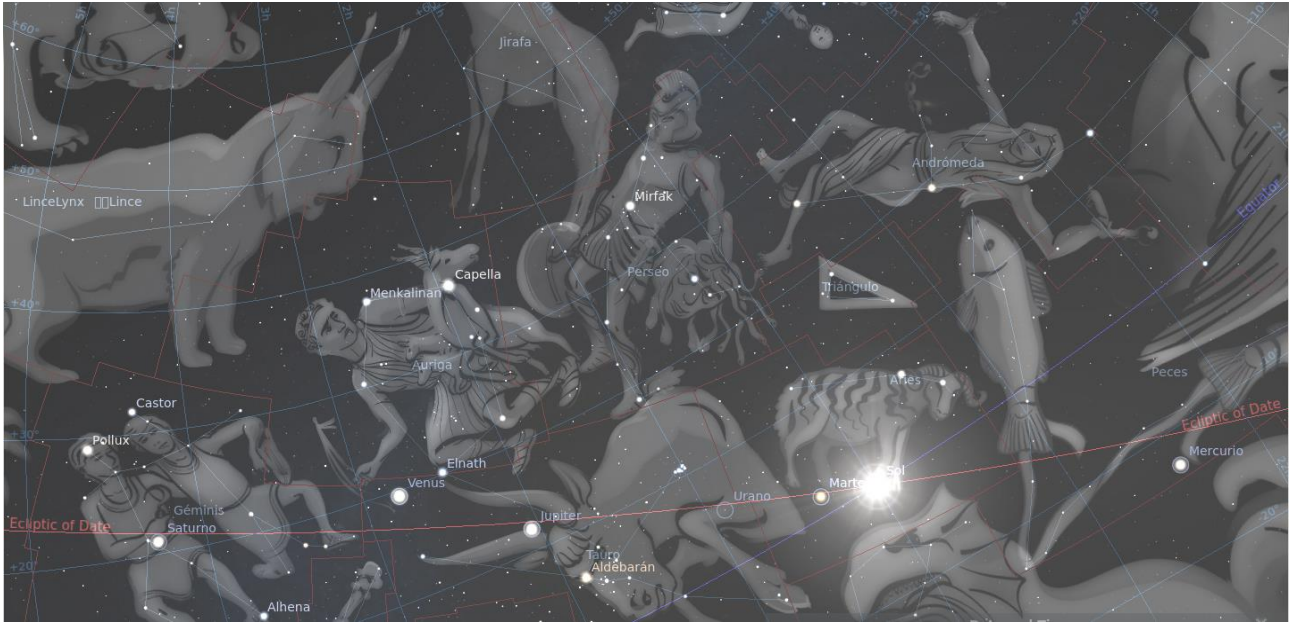

Fig. 9. Aspecto del cielo del equinoccio, visible desde la latitud de Roma en el I milenio a.C. Para esta época la "Era del Toro" ya había llegado a su fin (el punto vernal se había corrido hacia Aries). Por encima de Tauro se ve la imagen de Perseo con la cabeza de Medusa que lo distingue. Estos y otros elementos, son la base de la conjetura de que es Perseo quien toma el lugar de Mitra en la tauroctonía. Imagen del autor a partir de un mapa cortesía de *Stellarium*.

El héroe que sacrifica al toro simbolizaría entonces la fuerza cósmica que, en tiempos pasados, "destruyó el poder del toro al mover la totalidad de la estructura del cosmos de tal manera que el equinoccio de primavera se corrió fuera de la constelación de Tauro" (Ulansey, 1989: 83).

**Discusión**

En este trabajo hemos hecho un recorrido – parcial y sin duda incompleto – por la vida de una imagen que aún hoy presenta muchos enigmas. De los tres ejemplos célebres que Saxl decidió abordar en su estudio pionero: el ser divino que sujeta con sus manos a un par de serpientes que se vuelven hacia él, la imagen de la figura celestial alada de los albores de la civilización que se convirtió en el ángel de las sagradas escrituras y, finalmente, la lucha con la fiera que culmina con la dominación y el sacrificio del toro, este último es quizás el que tiene un anclaje más sólido con el cielo astronómico.

Como vimos, desde muy temprano se relacionó a la muerte simbólica del toro y a los demás personajes figurados en la entrada de la caverna – que aparecían en las composiciones usuales de los altares de los mitreos – con las constelaciones y con los grandes círculos de la esfera celeste (Chapman-Rietschi, 1997: 133). Pero uno de los principales misterios – en este culto de por sí mistérico – era no poder identificar al verdadero protagonista en el rol del Mitra matador de toros. Y varias fueron las ideas propuestas.

Mitra como el gigante Orión (Speidel, 1980), pues ya Porfirio, el neoplatónico del siglo III, en el capítulo 24 de su obra *De antro nympharum* (*El antro de las ninfas*), había escrito que Mitra estaba "ubicado sobre el ecuador [o círculo equinoccial], con el norte a su derecha y el sur a su izquierda" (North, 1990: 119; ver también Ramos, 1989; Campos Méndez, 2010: 65). Y Orión es justamente la constelación ecuatorial que lucha contra el toro. Otra propuesta fue no aislar a Mitra de su contexto iconográfico y enfocarse en la tauroctonía como un todo. Esta imagen sería entonces interpretada más como un reloj que como un calendario astronómico, que correlaciona trece características de la representación con constelaciones o partes de constelaciones que se ocultan en el horizonte occidental con intervalos de una hora aproximadamente (North, 1990: 125). Otra posibilidad, propuesta por Beck (2015: 1676), sugiere que la imagen representa a Mitra manifestándose en los cielos, no como un signo o constelación particular (como era el caso, por ejemplo, de Orión), sino cuando Sol se halla en la constelación de Leo.

Y la última en nuestra lista, aunque seguramente no entre las posibles interpretaciones de la imagen: Mitra como Perseo. Sin duda, la propuesta de Ulansey (1989: 26) tiene un cierto atractivo. En particular, la mayoría de las tauroctonías que se han descubierto hasta ahora muestran a Mitra que evita mirar al toro, exactamente en la manera en que se comporta Perseo quien, en la iconografía clásica, aleja su mirada de los ojos de la Medusa que acaba de matar, y así evita quedar convertido en piedra. Quizá sólo sea anecdótico, pero es notable que las cuatro imágenes de la tauroctonía (17, 18, 20 y 16.2) y las dos imágenes de los mitreos donde se alcanzan a ver los altares (15.4 y 16.1), que Warburg (2010: 29) presenta en el *Panel 8* de su *Bilderatlas Mnemosyne* (sobre la Ascensión hacia el Sol), todas tienen a Mitra mirando hacia otro lado [fig. 10].

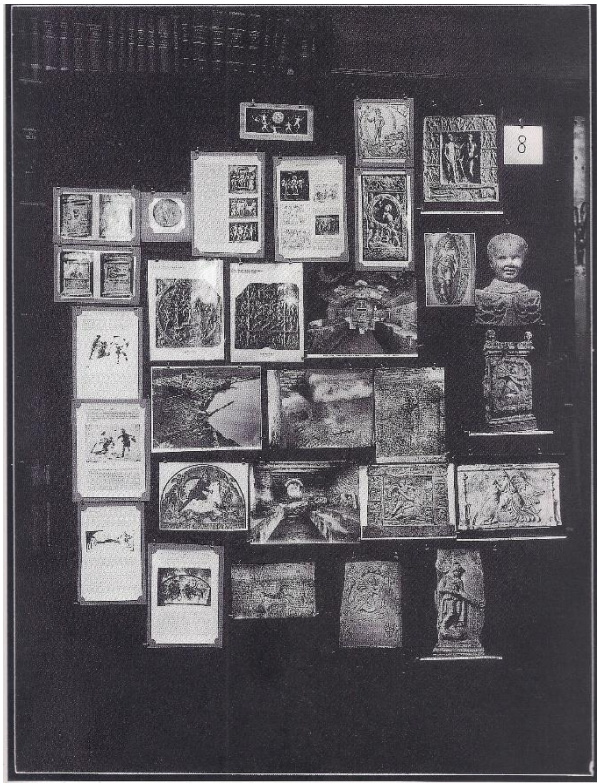 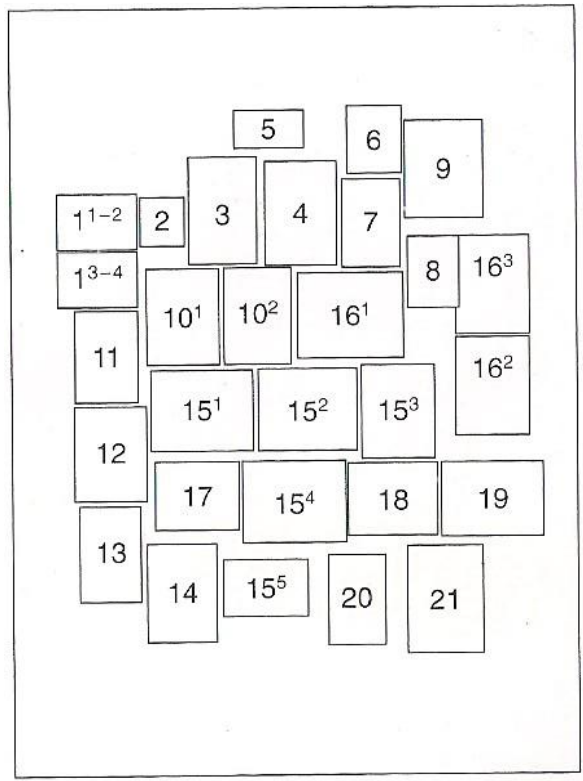

Fig. 10. Aby Warburg, *Panel 8* del *Bilderatlas Mnemosyne* y esquema numérico respectivo. (Warburg, 2010: 28-29)

Sin embargo, la conjetura de Ulansey de que un grupo de estoicos de la ciudad de Tarso, en algún momento después de mediados del siglo II a.C., convirtió las observaciones (e hipótesis) extremadamente técnicas de Hiparco sobre la precesión de los equinoccios en la doctrina fundacional del culto a Mitra, no es algo simple de digerir, y de hecho para la mayoría de los estudiosos resulta muy poco probable.

Se ha visto que, aparte de unos pocos autores tempranos – entre ellos, por supuesto, Ptolomeo, pero también el neoplatónico Proclo (siglo V), que niega su existencia, y Teón de Alejandría (siglo IV-V), quien repite los valores de Ptolomeo –, ninguno de los versados en astronomía mostró interés en la reconstrucción histórica de los equinoccios de épocas anteriores. De hecho, como señala Dreyer (1953: 203), la precesión parece no haber sido suficientemente conocida en la Antigüedad. Autores como Gémino de Rodas, Cleómedes, Teón de Esmirna, Manilio, Plinio el Viejo, Calcidio, Macrobio, o Marciano Capella (siglo V), entre otros, jamás aludieron a este gran hallazgo de Hiparco. En particular, Plinio (siglo I) tenía una gran debilidad por los hechos asombrosos y profesaba una gran admiración por Hiparco. De haber oído hablar sobre la precesión, seguramente lo habría reportado. Igualmente, Gémino (siglo I a.C.), quien era un escritor muy preciso en temas astronómicos y que estaba familiarizado con otros trabajos de Hiparco. Una breve mención sobre la precesión habría sido natural en su larga discusión sobre los signos zodiacales, si se hubiese enterado de ello, por supuesto (Evans, 1998: 262).

El hecho de imaginar que algún personaje de la Antigüedad pudo convertir esas sutilezas astronómicas en una religión es, en el decir de Beck (1998: 121), un gran anacronismo. Finalmente, otra crítica que puede hacerse a la conjetura de Ulansey es: ¿por qué este nuevo culto, que supuestamente se originó tan cerca del final de la Era de Aries (con el punto vernal en esa constelación) y en la víspera de la Era de Piscis (en la que estamos ahora), involucró tan sólo simbolismo arcaico relacionado con el toro?

Quizás estudios futuros, más profundos, puedan evidenciar mejor si la tauroctonía es warburguiana en algún aspecto. Y si las nociones antiguas de la doma o sacrificio de la bestia, representadas en sellos mesopotámicos y luego en otros soportes como vimos aquí, resurgen – o no – en un directo *Nachleben* en un nuevo marco cultural, romano y mistérico, de los primeros siglos de nuestra era. Es muy probable que, en esta búsqueda de darle renovada vida a lo antiguo, el cielo de las estrellas -la mitad del paisaje que nos acompaña desde la noche de los tiempos- haya jugado un rol no menor. Si así fuese, quizá las imágenes pinchadas en la tela oscura de los paneles del célebre Atlas puedan, en algunos casos, ser algún día complementadas con figuras del cielo, dibujadas con estrellas sobre el telón de fondo de la noche.

**Referencias**